\documentclass[prl,superscriptaddress,twocolumn,showpacs,preprintnumbers,amsmath,amssymb]{revtex4}

\usepackage{amsmath}    
\usepackage{graphicx}   
\usepackage{verbatim}   
\usepackage{color}      
\usepackage{subfigure}  
\usepackage{hyperref}   

\begin{document}
    \preprint{}
    \title{Superlens in the time domain}

    \author{Alexandre Archambault, Mondher Besbes, Jean-Jacques Greffet}
        \email[Corresponding author: ]{jean-jacques.greffet@institutoptique.fr}
    \affiliation{Laboratoire Charles Fabry, Institut d'Optique, Univ Paris Sud, CNRS, 2 av Fresnel,  91127 Palaiseau, France}

    \date{\today}

\begin{abstract}
It has been predicted theoretically and demonstrated experimentally that a planar slab supporting surface plasmons or surface phonon polaritons can behave as a super lens. However,  the resolution is limited by the losses of the slab. In this letter, we point out that the resolution limit imposed by losses can be overcome by using time-dependent illumination. 

\end{abstract}


    \maketitle

An important property of surface waves is their ability to focus the energy at subwavelength scales. This property is a consequence of the existence of surface waves with wavevectors $K$ parallel to the interface larger than $2\pi/\lambda$. For an air-metal interface, the well-known dispersion relation
$K=(\omega/c)\sqrt{\epsilon/(\epsilon+1)}$ shows that when $\epsilon$ approaches $-1$, the wavevector modulus diverges. The availability of arbitrary large wavevectors is the basis of the plasmonic super lens proposed by Pendry \cite{Pendry, Ramakrishna} and demonstrated by several groups \cite{Zhang, Taubner, Blaikie}. While experiments with subwavelength resolution have been reported, the improvement appears to be limited by the presence of losses as originally predicted \cite{Pendry} and analysed in several contributions \cite{Merlin,Smith, Pendry2}. Further developments have been focussed on other schemes such as the hyperlens \cite{hyperlens,hyperlens2}. In this letter, we show that using a time-dependent incident pulse allows improving the spatial resolution. 
We report a theoretical and numerical analysis of the imaging properties of a SiC planar slab using time-dependent illumination. We find that the resolution can be improved if the pulse duration is shorter or on the order of magnitude than the surface wave decay time.

At first glance, the idea of improving spatial resolution by using a time-dependent illuminating field seems dubious as a time-dependent field is a linear superposition of monochromatic fields, each of them being subject to a resolution limit due to losses. Let us first present two qualitative arguments suggesting that spatial resolution might indeed depend on the time shape of an incident pulse. It is well-known that the resolution limit can be analysed in terms of the exponential decay $\exp(-Kd)$ of the evanescent waves of a field at a distance $d$ from the object\cite{Novotnybook, ProgressinOptics}.  In his original proposal of a perfect lens, Pendry pointed out that in the electrostatic regime (for any large wavector $\mathbf{K}$), a silver slab with thickness $h$ has a p-polarized transmission connecting the amplitudes at both interfaces given by 
\begin{equation}
t(K,\omega)=\frac{4\epsilon\exp(-K h)}{(\epsilon+1)^2-(\epsilon-1)^2\exp(-2Kh)}.
\end{equation}
 The key point is that for $\epsilon(\omega)+1=0$, the transmission factor takes the value $\exp(Kh)$ so that the exponential decay $\exp(-Kh)$ appearing in the numerator is replaced by an enhancement factor $\exp(Kh)$ that compensates the decay of evanescent waves in the vacuum. It is useful to introduce the transmission factor $\tilde{t}(K,\omega)=t(K,\omega)\exp(-Kh)$ which takes value 1 for  $\epsilon(\omega)+1=0$. As discussed in the original paper and confirmed by latter analysis  and experiments, the presence of losses prevents to obtain  $\epsilon(\omega)=-1$ so that there is a practical limit to the resolution. 
 
 There is another point of view that is often used to explain the resolution limit due to losses. When plotting the dispersion relation of a surface plasmon propagating along a metal-vacuum interface,  assuming that the frequency is real, the relation $K=(\omega/c)\sqrt{\epsilon/(\epsilon+1)}$ yields a complex wavevector. When plotting the dispersion relation in the plane ($Re(K),\omega$), a backbending is observed (see Fig.1). Hence, the dispersion relation shows a maximum value of the wavector $K_{max}$ corresponding to the turning point. From Fourier analysis, a resolution limit $\Delta x=\pi/ K_{max}$ is expected.

 The point that we want to make here is that these arguments are implicitly assuming that the measurement is performed in stationary regime with a harmonic field. Let us revisit these arguments allowing $\omega$ to be a complex frequency. Equation $\epsilon(\omega)+1=0$ can now be solved exactly if we seek a complex solution $\omega_{sw}$. Of course, a wave varying as $\exp(-i\omega_{sw} t)$ with $Im(\omega_{sw})<0$ diverges for negative times so that it is not an acceptable solution for all times. However, we may consider approaching this function in a limited time window. Improving the resolution will depend  on our ability to approach this ideal solution. Obviously, such a solution has a time-dependent amplitude which indicates that \textit{a time-dependent illumination has to be used}. Finally, this first remark also provides the relevant time scale for the time-dependent illumination: the surface wave decay time $1/Im(\omega_{sw})$.

Let us now revisit the discussion of the dispersion relation for lossy materials. While solving the equation $K=(\omega/c)\sqrt{\epsilon/(\epsilon+1)}$ with a real frequency and a complex $K$ yields a dispersion relation with a backbending, it turns out that by solving the same equation with a complex frequency and a real wavevector $K$ yields a dispersion relation without backbending as shown in Fig.1. Hence, the dispersion relation for complex frequencies displays an asymptote \textit{with arbitrary large wavevectors}. There is a price to pay. The surface waves with very large wavevectors which may contribute to a highly localized field are very short-lived modes. Note also that they have an almost zero group velocity. This is not a drawback for the superlens imaging process which is essentially an electrostatic image. For plasmons, the typical time scale is on the order of 10 fs whereas for surface phonon polaritons, decay times are on the order of 1 ps. Again, we see that it is possible to recover large wavevectors and therefore improve the resolution by working with time-dependent fields.  Let us note in passing that the dispersion relation with an asymptote not only leads to the prediction of large spatial resolution but also to large electromagnetic density of states close to the interface. This large density of states can be independently calculated in the direct space and has a divergence \cite{Ford}. Two well-known physical consequences of this divergence are the extremly short decay time of a two-level system close to an interface \cite{Chance} and the divergence of the thermal radiation density of energy close to the interface\cite{Shchegrov}. The existence of two dispersion relations which can both being meaningful depending on the context was recognized in the early days of surface plasmons \cite{Alexander}. A detailed discussion can be found in ref.\cite{Archambault}.

\begin{figure}[!htbp]
\centering
\includegraphics[width=80mm]{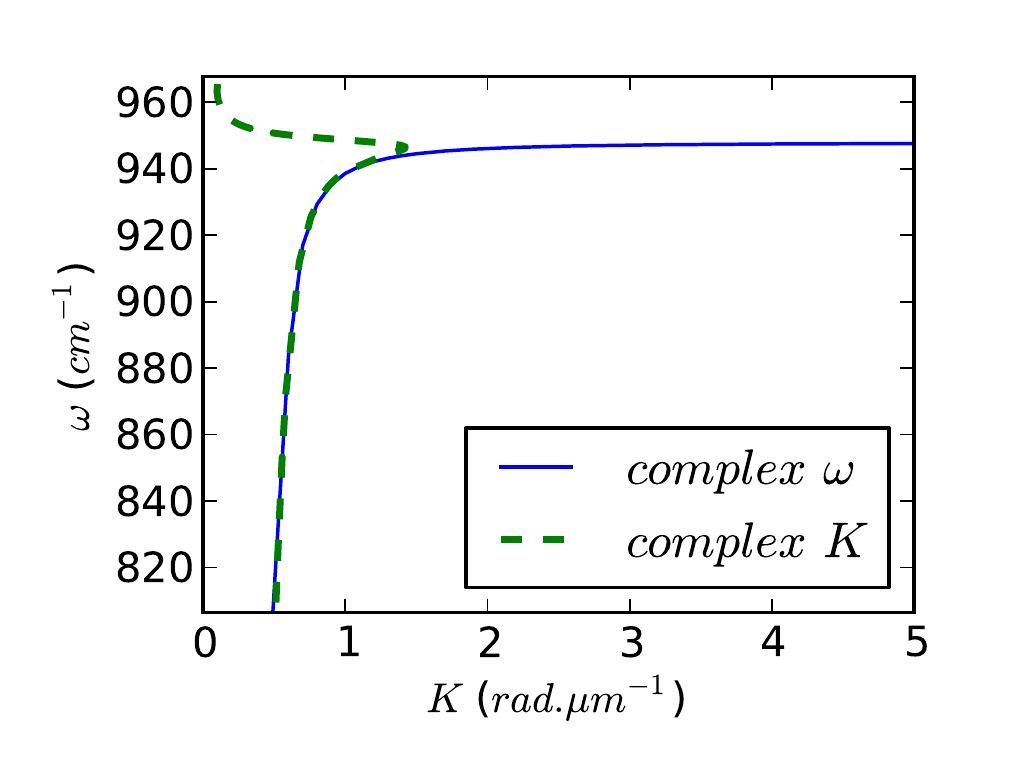}
\caption{Dispersion relation of a surface phonon polariton at the interface SiC/air displaying the real part of $\omega$ versus the real part of $K$. Green/dashed :  the wavector is chosen to be complex, Blue/plain : the circular frequency is chosen to be complex. } 
\label{FIG1}
\end{figure}

From the previous discussion, we are led to revisit the resolution of a super lens consisting of a thin symmetric slab supporting surface waves. A number of conditions need to be fulfilled in order to achieve super-resolution:  i) The carrier frequency has to be chosen close to the surface wave frequency $\omega_{sw}$ satisfying $\epsilon(\omega_{sw})+1=0$ in order to ensure an efficient excitation of the surface waves with large wavevector ii) the contribution of the transient surface wave needs to be the leading contribution to the field. If a stationary field builds up, it may overwhelm the transient contribution. Since the transient surface waves decay exponentially with a typical time scale given by $1/Im(\omega_{sw})$, we can separate them from the incident field using a shorter pulse so that the resonantly excited surface waves are the only contributor to the near field after the end of the incident pulse. This amounts to perform a \textit{time filtering} of the large $\mathbf{K}$ which contribute to the large resolution. This is the major difference with previous work that considered a time analysis of the transient regime \cite{GomezSantos,Pendry2} or a relatively long pulse illumination\cite{Kik}. For plasmons, the second condition implies pulses shorter than 10 fs. For surface phonon polaritons, pulses duration need to be on the order of  a picosecond. 

We now show a numerical simulation of the image of a time-dependent point-like dipole $\mathbf{p} f(t)\delta(\mathbf{r}-\mathbf{r}')$. Since we are working in the non-retarded approximation, the field spatial structure can be derived using electrostatics. We derive the potential $\phi(\mathbf{r},\mathbf{r}_0,t)$ created at $\mathbf{r}$ by a time-dependent charge $ q_0\delta(t-t')$ located at $\mathbf{r}_0$. The potential $\phi_{\mathbf{p}}(\mathbf{r},\mathbf{r}_0,t)$ produced by a dipole $\mathbf{p}$ can then be derived using $\mathbf{p}\cdot\nabla_{\mathbf{r}_0}\frac{\phi(\mathbf{r},\mathbf{r}_0,t)}{q_0}$. The electric field is then given by $\mathbf{E}(\mathbf{r},t)=-\nabla_{\mathbf{r}}\phi_{\mathbf{p}}(\mathbf{r},\mathbf{r}_0,t)$. The system studied here is a SiC slab with thickness $d$ in a vacuum as depicted in Fig.2. The dielectric constant is given by 
$\epsilon(\omega)=\epsilon_{\infty}\frac{\omega_{LO}-\omega^2-i\Gamma\omega}{\omega_{TO}-\omega^2-i\Gamma\omega},$  where $\epsilon_{\infty}=6.7$, $\omega_{TO}=793$ cm$^{-1}$, $\omega_{LO}=969$ cm$^{-1}$, and $\Gamma=4.76$ cm$^{-1}$. For this material, the condition $\epsilon(\omega)+1=0$ is satisfied for $\omega_{sw}$ with $Re(\omega_{sw})=947.7$cm$^{-1}$. The corresponding decay time is $1/2 Im(\omega_{sw})= 1.1$ps. The potential produced by a harmonic unit charge at $\mathbf{r}=(x,y,z)$ with $\rho=\sqrt{x^2+y^2}$ in a vacuum is given by:
\begin{eqnarray}
\phi(\mathbf{r},\omega)&=&\frac{q_0 f(\omega)}{4\pi\epsilon_0}\int_0^{\infty}\mathrm{d} K J_0(K \rho)\exp(-K [z-(z_0+2h)]) \nonumber \\
&=&
\frac{q_0 f(\omega)}{4\pi\epsilon_0}\frac{1}{\sqrt{\rho^2+[z-(z_0+2h)]^2}}. 
\end{eqnarray}

\begin{figure}[!htbp]
\centering
\includegraphics[width=80mm]{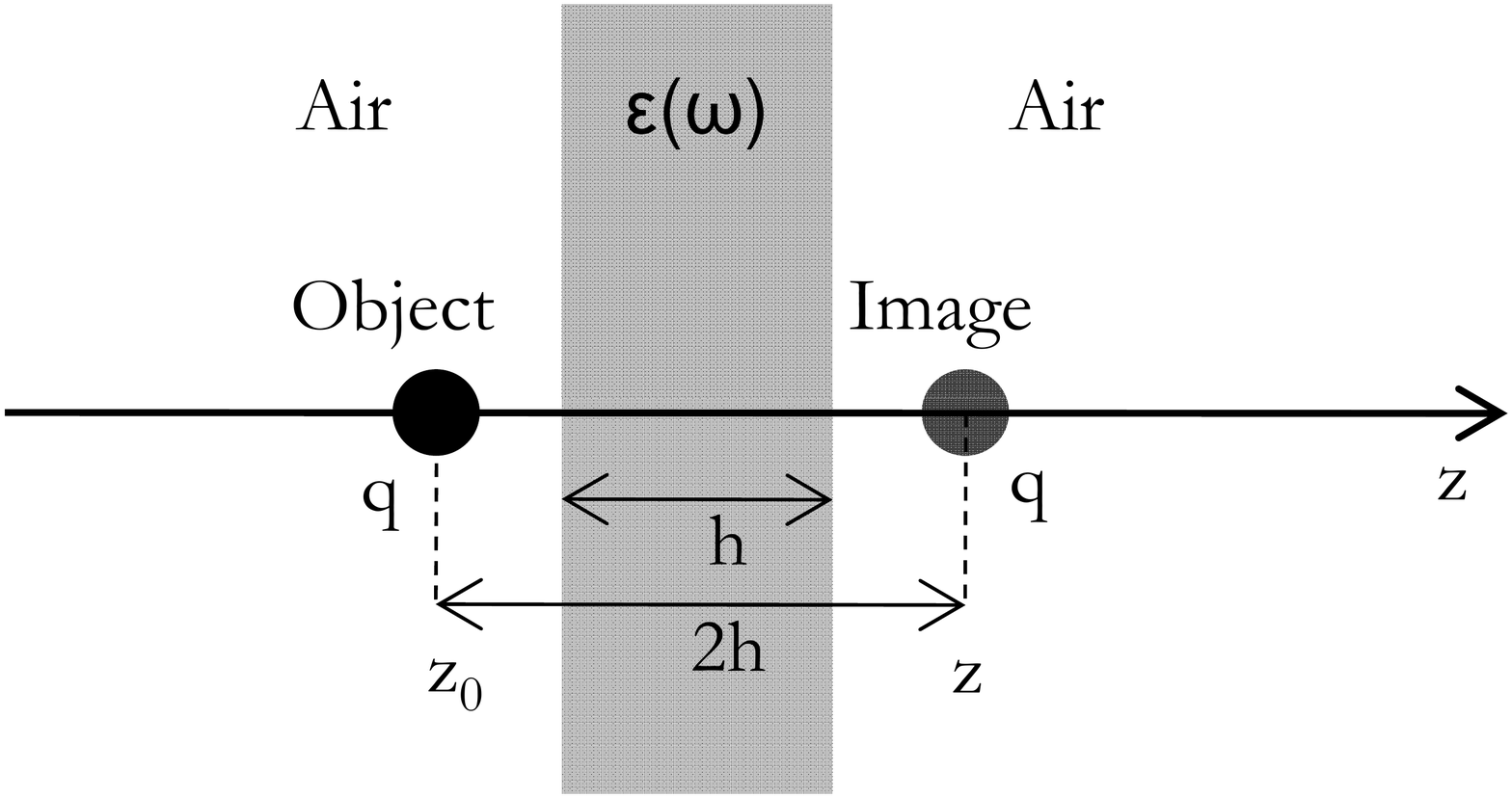}
\caption{Scheme of the planar lens with thickness $h$. The image is located at $z=z_0+2h$. } 
\label{FIG2}
\end{figure}

In the presence of a slab, the potential produced on the other side of the slab is found by introducing the frequency-dependent transmission factor:

\begin{eqnarray}
\phi(\mathbf{r},\omega)&=\frac{q_0 f(\omega)}{4\pi\epsilon_0}\int_0^{\infty}\mathrm{d} K \, \tilde{t}(K, \omega) \,J_0(K \rho) \nonumber \\&\exp(-K [z-(z_0+2h)]). 
\end{eqnarray}

The value of this filter is plotted in Fig.3 for three different frequencies as a function of the wavector $K$. For the complex frequency  $\omega_{sw}$, the filter does not depend on $K$. For real frequencies, the filter is band limited. Out of resonance, the filter decays very fast. As expected, the cut-off frequency is larger for  $Re(\omega_{sw})=947.7 cm^{-1}$.
\begin{figure}[!htbp]
\centering
\includegraphics[width=80mm]{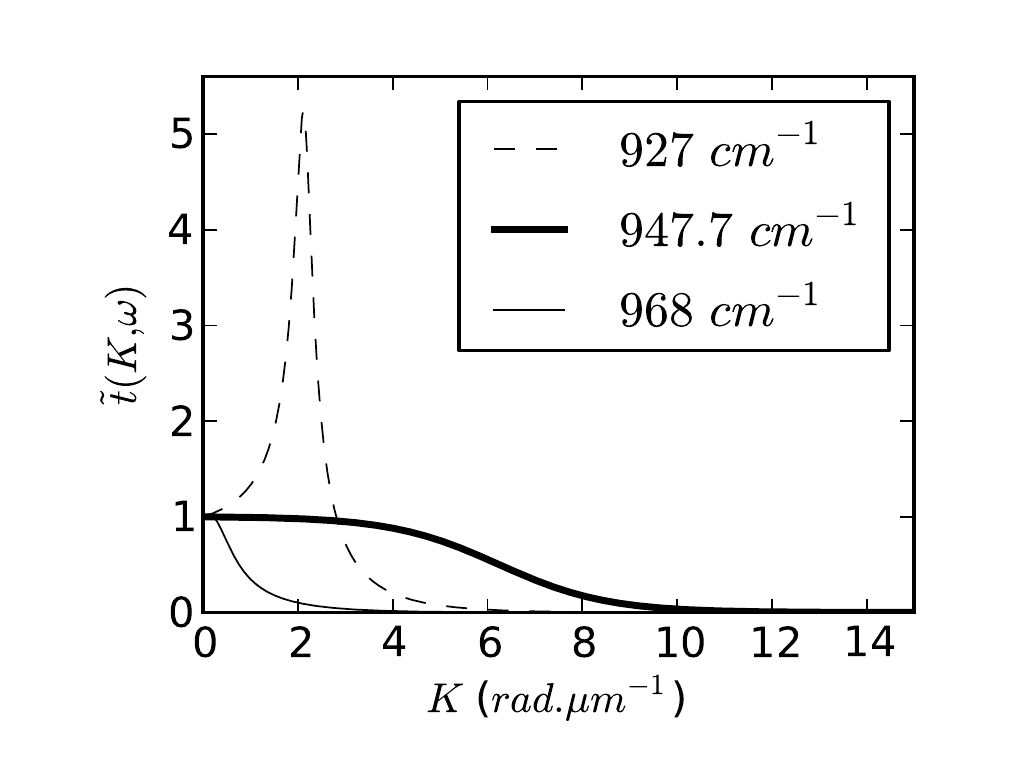}
\caption{Transmission factor as a function of the wavevector for three different frequencies. An ideal superlens is characterized by a transmission factor $\tilde{t}(K, \omega)=1$ at the optimum frequency. In the presence of losses, we obtain the best behaviour at $\omega=Re(\omega_{sw})$.} 
\label{FIG2bis}
\end{figure}

In practice, we use a time-dependent signal composed of real frequencies. Hence the key issue is the design of the best possible time shape of the pulse. Here, we have not attempted to optimize the pulse shape. For the sake of illustration, in what follows, we consider a stationary monochromatic field which is turned off at $t=0$. The surface wave field then naturally decays with a frequency close to $\omega_{sw}$ and we study the field for $t>0$. The carrier frequency is taken to be the real part of the surface wave resonant frequency $\omega'_{sw}=Re[\omega_{sw}]$ for a single interface so that the incident field is given by $f(t)=H(-t)\exp(-i\omega'_{sw}t)$. 

\begin{figure}[!htbp]
\centering
\includegraphics[width=80mm]{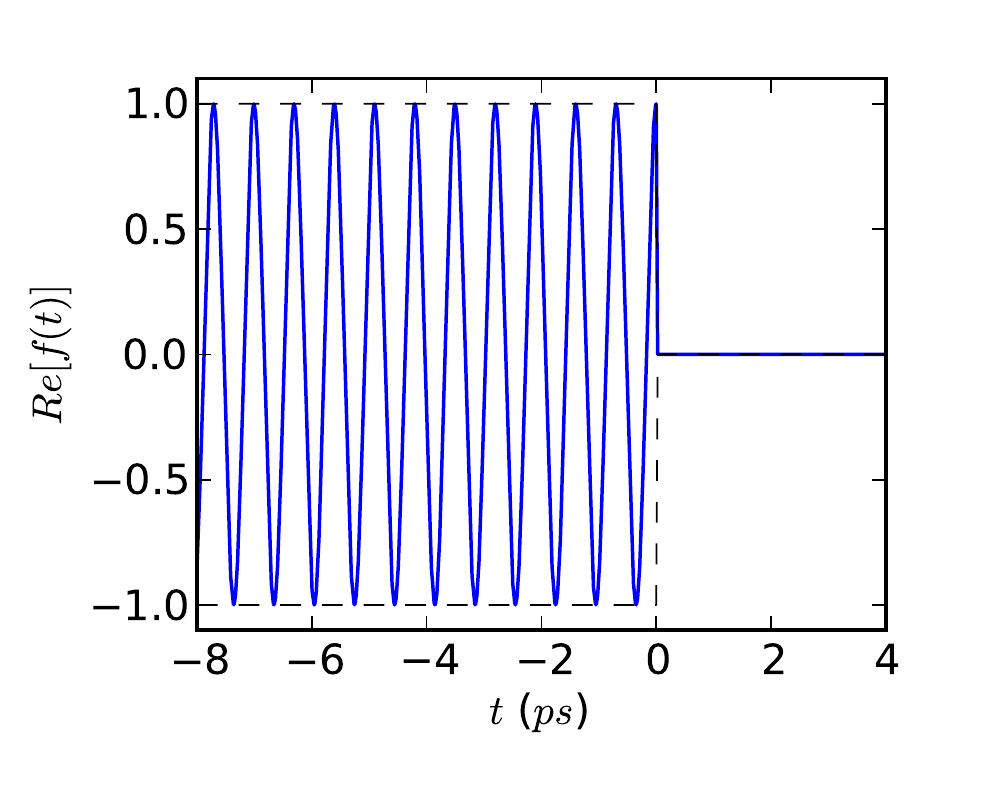}
\caption{Incident field $H(-t)\exp(-i\omega'_{sw}t)$ used to excite the system mode with complex frequency.} 
\label{FIG3}
\end{figure}

We are now ready to evaluate the field in time domain. The potential is given by:
\begin{equation}
\phi(\mathbf{r},t)=\int_{-\infty}^{\infty}\frac{\mathrm{d}\omega}{2\pi}\,\phi(\mathbf{r},\omega) \exp(-i\omega t)
\end{equation}

 The results are shown in Fig.5. We consider a dipole oriented perpendicularly to the interface. We plot the square of the z-component of the electric field averaged over a few cycles. The intensity is shown at different times $t>0$. In order to compare the resolution at different times $t>0$, we have compensated the exponential decay of the amplitude. It is clearly seen that the width at half-maximum of the intensity distribution decreases as time increases showing that the resolution is increased. It is seen that for a 16 ps delay, the width is reduced to $60\%$ of its value for a monochromatic illumination. Of course, the price to pay for this resolution enhancement is the exponential decay of the amplitude. Note however that the incident field is extinguished so that detecting the signal is only limited by the signal to noise ratio.  

\begin{figure}[!htbp]
\centering
\includegraphics[width=80mm]{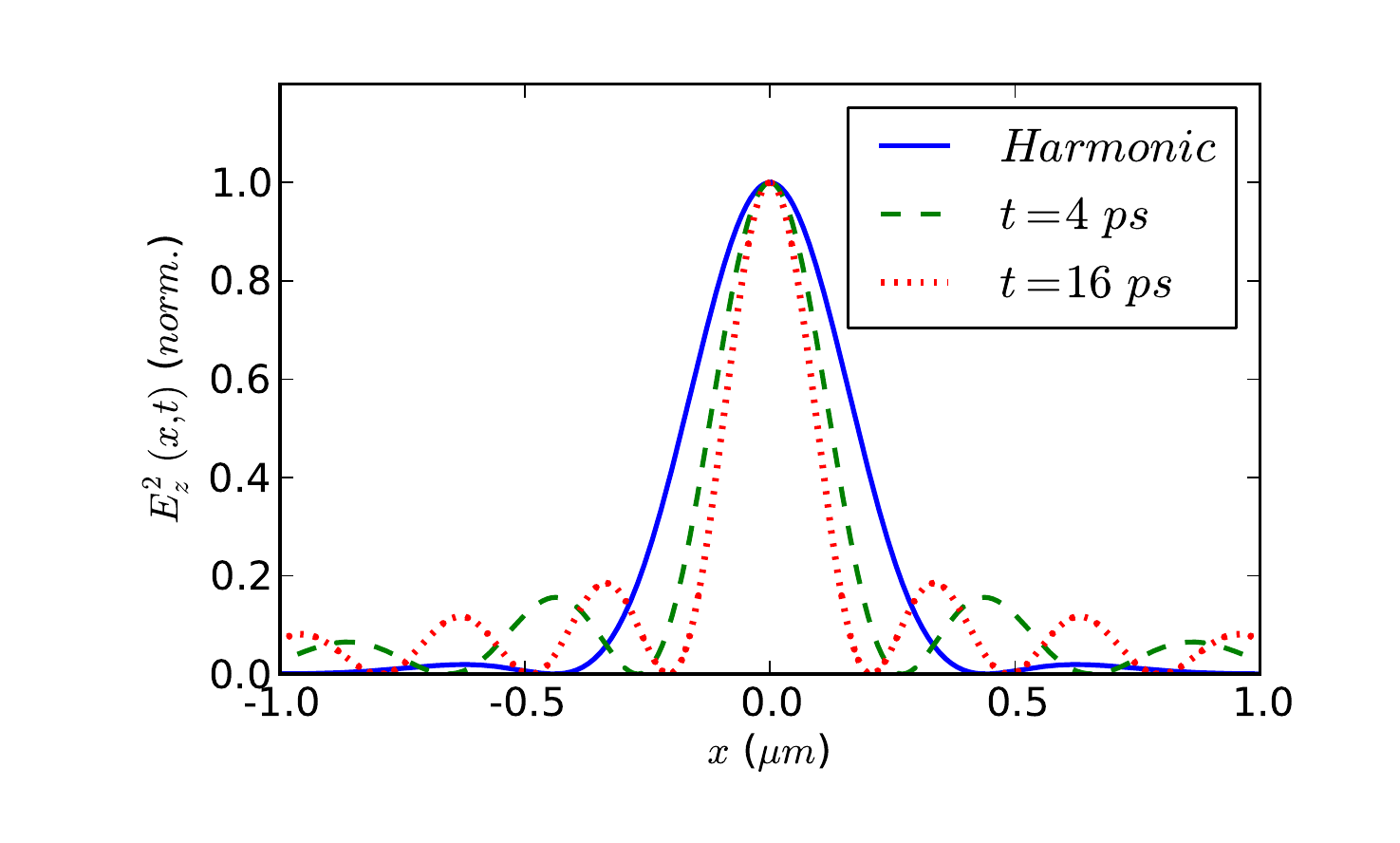}
\caption{Normalized profile of the intensity at the image location.  A 60\% reduction of the width of the pulse is observed at $t=16$ps. The amplitudes habe been normalized in order to compare the widths.} 
\label{FIG5}
\end{figure}

So far, we have introduced the idea that time-dependent illumination allows exciting surface waves with no limitation regarding the wavevector. Yet, we have not analysed the detailed mechanism. Analysing the behaviour of the planar lens requires to examine the image formation in terms of spatial frequencies filtering in the time domain. To proceed, we assume that the time dependent source $q_0 f(t)$ has been induced by an incident electric field $\mathbf{E}(t)=\mathbf{E}_0 f(t)$. We introduce the frequency spectrum of the time dependent charge $q(t)=q_0\int f(\omega)\exp(-i\omega t) \mathrm{d}\omega/2\pi$. The time-dependent potential transmitted by the superlens is thus given by :
\begin{eqnarray}
\phi(\mathbf{r},t)&=&\frac{q_0}{4\pi\epsilon_0}\int_0^{\infty}\mathrm{d} K J_0(K \rho)\exp(-K [z-(z_0+2h)]) \nonumber \\
&\times &\int_{-\infty}^{\infty}\tilde{t}(K ,\omega)f(\omega)\exp(-i\omega t)\frac{\mathrm{d}\omega}{2\pi}. 
\end{eqnarray}

The last integral appears to be a \textit{time-dependent filter} $\Pi(K, t)$. Hence, a proper choice of $f(\omega)$ may allow tailoring the spatial frequency filter. The filter $\Pi(K, t)$ can be evaluated analytically for a given pulse. The transmission factor $\tilde{t}(K, \omega)$ of a symmetric slab has poles associated with the symmetric and antisymmetric surface modes. Their excitation generates two contributions oscillating at the surface waves frequencies. A beating is observed as already discussed in refs \cite{GomezSantos,Pendry2}. Here, we see that the choice of the carrier frequency and the enveloppe of the pulse provides an additional degree of freedom to control the modes excitation. The structure of the filter is shown in Fig.6 for the step function illumination. It is seen that the filter has a broader bandwidth than the stationary filter at $\omega_{sw}$. The closed-form expression can be found in ref.\cite{these}.

\begin{figure}[!htbp]
\centering
\includegraphics[width=80mm]{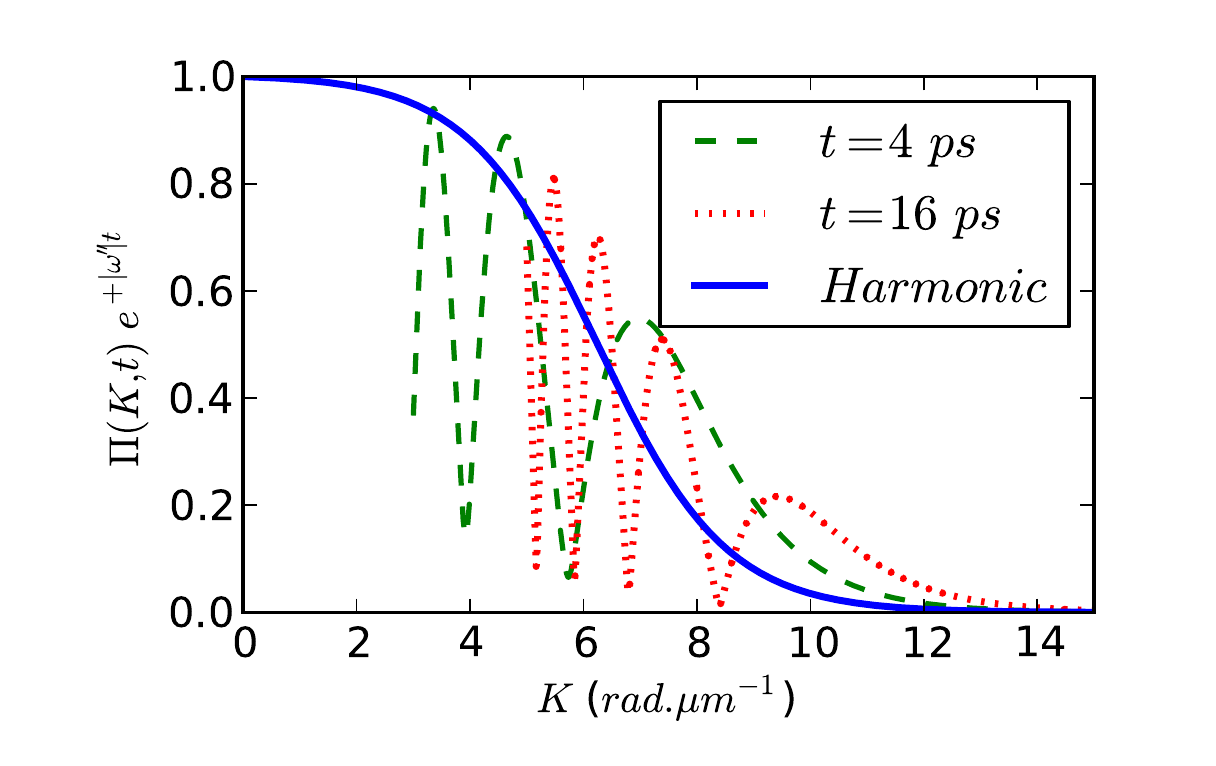}
\caption{Spatial frequency filter $\Pi(K, t)$ at different times. It is seen that the bandwidth increases. This filter depends on the choice of the incident field. } 
\label{FIG6}
\end{figure}

 In summary, we have put forward the idea that spatial resolution using lossy surface waves can be improved by using time-dependent illumination. The fundamental idea illustrated here is that when dealing with time-dependent fields, it is relevant to think in term of modes with real wavevector and complex frequency (i.e. limited lifetime). The associated dispersion relation does not present any spatial frequency cut-off. A direct simulation confirms these ideas and shows a 60\% reduction of the width of the image of a point source. Further optimization of the time shape of the incident pulse should allow improving further the resolution. Yet, the general trend is that the increase of the spatial bandwidth comes at the cost of an exponential decay of the amplitude of the field. Hence, although the resolution is not theoretically limited with this time-dependent scheme, the signal to noise ratio introduces a severe limitation. 



\begin{references}

\bibitem{Pendry} J.B. Pendry, Phys.Rev.Lett. \textbf{85}, 3966 (2000).
\bibitem{Ramakrishna} J. Pendry, S. Ramakrishna, Physica B: Condensed Matter \textbf{38} 329 (2003).
\bibitem{Zhang} N. Fang, H. Lee, C. Sun, X. Zhang, Science \textbf{308}, 5721 (2005).
\bibitem{Taubner} T. Taubner, D. Korobkin, Y. Urzhumov, G. Shvets, R. Hillenbrand, Science \textbf{313}, 1595 (2006).
\bibitem{Blaikie} D.O.S. Melville, R.J. Blaikie, C.R. Wolf, Appl.Phys.Lett. \textbf{84} 4403 (2004).
\bibitem{Merlin} R. Merlin, Appl.Phys.Lett. \textbf{84}, 1290 (2004).
\bibitem{Smith}  D. R. Smith, D. Schuring, M. Rosenbluth, S. Schultz, S. A. Ramakrishna, J. B. Pendry, Appl. Phys. Lett. \textbf{82}, 1506 (2003).
\bibitem{Pendry2} W.H. Wee, J.B. Pendry, Phys.Rev.Lett. \textbf{106}, 165503 (2011).
\bibitem{hyperlens}  Z.Jacob,  L. V. Alekseyev, E. Narimanov,  Opt. Express \textbf{14}, 8247 (2006).
\bibitem{hyperlens2} Z. Liu et al. Science \textbf{315}, 1686 (2007)
\bibitem{Novotnybook}  B. Hecht, L. Novotny, Near-field Optics, (Cambridge University Press, Cambridge, 2003).
\bibitem{ProgressinOptics} J.J. Greffet, R. Carminati, Image formation in Near-field optics, Progress in Surface Science \textbf{56},  133(1997).
\bibitem{Ford} G.W.Ford, W.H.Weber, Surf.Sci. \textbf{129}, 123 (1983).
\bibitem{Chance} R.R. Chance, A. Prock, R. Silbey, Adv. Chem. Phys. \textbf{37}, 1 (1978).
\bibitem{Shchegrov} A.V. Shchegrov, K. Joulain, R. Carminati, J.J. Greffet, Phys. Rev.Lett. \textbf{85} 1548 (2000)
\bibitem{Alexander} R. W. Alexander, G. S. Kovener, and R. J. Bell, Phys. Rev. Lett. \textbf{32}, 154 (1974).
\bibitem{Archambault} A. Archambault, T.V. Teperik, F. Marquier, J.J. Greffet, Phys.Rev.B \textbf{79}, 195414 (2009).
\bibitem{GomezSantos} G. Gomez-Santos, Phys.Rev.Lett. \textbf{90}, 077401 (2003).
\bibitem{Kik} P.G. Kik, S.A. Maier, H. A. Atwater, Phys.Rev.B \textbf{69}, 045418 (2004).
\bibitem{these} A. Archambault, Optique des ondes de surface : super-rŽsolution et interaction matiere-rayonnement, Thesis Universit\'e Paris Sud, (2011)URL:http://tel.archives-ouvertes.fr/tel-00678073/fr


\end{references}
\end{document}